\documentclass[preprint,showpacs,preprintnumbers,amsmath,amssymb]{revtex4}
\usepackage{graphicx}

\begin{document}

\title{One dimensional quantum walk with unitary noise}

\author{Daniel Shapira}
\author{Ofer Biham}
\affiliation{Racah Institute of Physics, 
The Hebrew University, Jerusalem,91904,Israel}

\author{A.J. Bracken}
\author{Michelle Hackett}
\affiliation{Department of Mathematics, University of Queensland,
Brisbane, Queensland 4072, Australia}
\newpage

\begin{abstract}
The effect of unitary noise on the discrete one-dimensional quantum walk 
is studied using computer simulations. 
For the noiseless quantum walk, starting at the origin
($n=0$) at time $t=0$, the position distribution
$P_t(n)$ 
at time $t$ is very different from 
the Gaussian distribution obtained for
the classical random walk.
Furthermore, its standard deviation, $\sigma(t)$
scales as
$\sigma(t) \sim t$,
unlike the classical random walk for which
$\sigma(t) \sim \sqrt{t}$.
It is shown that when the quantum walk is exposed to unitary noise, 
it exhibits a crossover from quantum behavior for short times to 
classical-like behavior for long times. The crossover time is found to be 
$T \sim {\alpha}^{-2}$ where $\alpha$ is the standard deviation of 
the noise. 
\end{abstract}


\maketitle

\section{Introduction}
Random walk models describe a great variety of diffusion phenomena 
in physical systems. Such phenomena include the diffusion of particles 
in a fluid (Brownian motion), the motion of vacancies in a crystal and 
of atoms on a crystalline surface. 
Related models are also used to describe the spatial structure of 
systems such as polymer chains.

A random walker on a lattice hops at each time step from its present site 
to one of its nearest neighbors. The hopping direction is picked 
randomly,  
with, for example,
the probability to hop to each of $c$ nearest neighbors 
given by $1/c$.
Random walk models on discrete lattices 
as well as in the continuum have been studied extensively. 

Consider a random walker 
on a one dimensional lattice. 
Denote the probability to find it at site $n$ at time 
$t$ by 
$P_t(n)$. 
The time evolution of 
$P_t(n)$
is described by the recursion equation

\begin{equation}
P_{t+1}(n) = \frac{1}{2} [ P_t(n-1) + P_t(n+1) ], \ \ n=0,\pm 1, \pm 2,\dots
\label{eq:recursionP}
\end{equation}

\noindent
For a walker starting from the origin at 
$t=0$,
the probabilities,
$P_t(n)$, of the walker to be at site $n$ at time $t$,
are given by the components of the Pascal triangle,
namely

\begin{equation}
P_t(n) = 
\left\{
\begin{array}{ll}
\frac{t!}{2^t \left(\frac{t-n}{2} \right)! \left(\frac{t+n}{2} \right)!}:
& \ \   n=-t,-t+2,\dots,t 
\\
0: 
&  n=-t+1,-t+3,\dots,t-1,
\end{array}
\right.
\label{eq:Pascal}
\end{equation}

\noindent
and 
$P_t(n)=0$
for 
$|n| > t$.
Note that at even times only the even sites can be
occupied, while at odd times only the odd sites can
be occupied.
A continuum description of the random walk can be obtained
when the lattice constant
$\Delta n \rightarrow 0$
and the time step
$\Delta t \rightarrow 0$,
such that
$D = \Delta n^2/(2 \Delta t)$
converges to a finite value.
The parameter $D$ is called the diffusion
coefficient and it quantifies the rate in which the
random walker moves.
In this limit the random walk can be described by the
diffusion equation

\begin{equation}
\frac{d P_t(n)}{dt} - D \frac{d^2 P_t(n)}{dn^2} = 0,
\label{eq:diffusion}
\end{equation}

\noindent
where $n$ and $t$ are continuous variables.
The solution of this equation,
approached by Eq. 
(\ref{eq:Pascal})
in this limit,
is the Gaussian probability
distribution

\begin{equation}
P_t(n) = \frac{1}{\sqrt{4 \pi D t} } 
e^{ - \frac{n^2}{4 D t} } 
\label{eq:Gaussian}
\end{equation}

\noindent
with $D=1/2$.
The standard deviation 
of this distribution,
$\sigma(t) = \sqrt{2 D t}$,
thus takes the form
$\sigma(t) = \sqrt{t}$.

Recently, quantum analogues of the classical 
random walk model have been studied 
on the one dimensional lattice 
\cite{AharonovY93,Meyer96,Ambainis01,Mackay01,Travaglione2002}
as well as on more general graphs 
\cite{AharonovD01,Childs01}. 
Consider the quantum walk on the one dimensional 
lattice. 
Unlike the classical random walker that occupies a single site at
a time, the quantum walker can be in an extended state. 
This state is
a superposition of 
all the basis states
$| n \rangle$, 
$n=0,\pm 1,\pm 2,\dots$,
where 
$| n \rangle$  
is the state in which the walker is
located at site $n$.
In addition to the spatial degree of freedom 
the walker has a chirality qubit, which 
can be in a superposition of the states
$| R \rangle$
and 
$| L \rangle$,
and
determines the 
direction of the next hop.
Each move of the walker consists of two unitary operations.
The first one is 
a unitary transform, taken here to be
the Hadamard transform, on the chirality
qubit (expressed in the standard basis).
The second one is the actual move, in which a walker with
chirality
$| R \rangle$
moves to the right and 
a walker with chirality
$| L \rangle$
moves to the left.
This definition resembles the classical walk in the sense
that a walker in a basis state $| n \rangle | s \rangle$
($s=R$ or $L$) at time $t$,
will have equal probabilities to be found in the $n+1$
and $n-1$ sites if a measurement is performed at time
$t+1$.
However, the coherent motion of the state gives rise to
constructive and destructive interference that strongly
modifies the emerging probability distribution. 
As a result, a quantum walk 
starting from the origin at time $t=0$
moves away faster than 
the classical random walk. 
The standard deviation 
$\sigma(t)$
of the probability
distribution
$P_t(n)$ 
for the quantum walk
increases with time
according to
$\sigma(t) \sim t$,
while for the classical random walk
$\sigma(t) \sim \sqrt{t}$.

One of the main difficulties in the experimental implementation 
of quantum algorithms is the sensitivity of the quantum systems to 
noise 
\cite{Bernstein97,Preskill,Nielsen00} 
and decoherence  
\cite{Zurek01}. 
The problem of decoherence can be tackled by
using quantum error correction methods 
\cite{Shor95,Shor96a,Steane96,Steane96a,Knill97} 
as well as 
decoherence free sub-spaces  
\cite{Zanardi97,Lidar98,Bacon99,Kempe01}.
However, these methods require significant overhead,
making the quantum circuits more complicated. 
It is thus useful to examine the effect of decoherence on various
quantum algorithms, implemented on unprotected quantum circuits
\cite{Azuma2002}.
The quantum walk model is particularly suitable for this task
since its behavior is qualitatively different from the classical
random walk.
The effect of decoherence on the quantum walk has been studied
recently
\cite{Brun2002,Kendon2002,Brun2003,Kendon2003,Kendon2003b}.
It was found that the quantum walk is highly sensitive to decoherence, 
namely, even weak decoherence gives rise to classical-like behavior
in the long time limit.
Decoherence is a fundamental problem because it is
a result of the unavoidable interaction of the system with the environment,
that makes the two of them entangled.
This interaction can be described by a variety of noise models, 
bringing the system into a mixed state, by essentially tracing out the
environment degrees of freedom.

A related problem is the effect of
unitary noise, which appears as a 
result of fluctuations and drifts in the generating Hamiltonian of 
any given unitary operation
\cite{Shapira2003}.
This noise tends to reduce the performance of quantum devices,
but does not cause an entanglement with the environment, namely
the system remains in a pure state.

In this paper we analyze the effect of unitary noise on the discrete 
one-dimensional quantum walk model, using computer simulations. 
We find that even a tiny noise level will eventually induce a crossover 
from the quantum walk into a
classical-like behavior, characterized by
$\sigma(t) \sim \sqrt{t}$. 
The crossover time $T$ is calculated numerically.
It is found that 
$T \sim \alpha^{-2}$, 
where $\alpha$ 
is the standard deviation of 
the noise distribution. 

The paper is organized as follows.
The quantum walk model on the one-dimensional 
lattice is described in Sec. II. 
The noisy quantum walk is introduced in Sec. III. 
Simulations and results are presented in Sec. IV, 
followed by a summary in Sec. V.  

\section{The quantum walk model}

Consider a quantum walker on a one-dimensional lattice. 
The state of the system at time $t$ is
 
\begin{equation}
| \psi(t) \rangle = 
\sum_{n=-\infty}^{\infty} \sum_{s=L}^R
a_{n,s} (t) | n \rangle | s \rangle, 
\label{eq:psi_t}
\end{equation}

\noindent
where the amplitudes  
$a_{n,s}$ 
are complex numbers that satisfy 
$\sum_{n,s} {|a_{n,s}|}^2 = 1$. 
The state vector 
$| n \rangle$, 
where 
$n = 0, \pm 1, \pm 2 , \ldots$, 
represents the position of the walker 
on the lattice. 
The state vector 
$|s\rangle = |R\rangle$ 
or $|L\rangle$ 
represents the chirality degree of freedom. 
The chirality consists of a single qubit, and its state 
determines 
the coefficients
$a_{n,s}(t+1)$.
Operations in the chirality space in the quantum model, 
replace the randomized decision that appears 
in the classical walk model and determines the hopping direction.  
When measuring the walker position 
at a certain time $t$, the probability to find the walker at site  
$n$ is given by:

\begin{equation}
P_t (n) = {|a_{n,R}(t)|}^2 + {|a_{n,L}(t)|}^2.
\label{eq:Pnt}
\end{equation}

\noindent
The time evolution of the quantum walk is expressed by

\begin{equation}
| \psi(t+1) \rangle  =  \hat{Q}_0  | \psi(t) \rangle,   
\label{eq:psi_of_walk}
\end{equation}         

\noindent
where $| \psi(t) \rangle$ is the quantum state at time $t$. 
The quantum walker's step is defined by  

\begin{equation}
\hat{Q}_0 = \hat{T} \hat{U}_0,
\label{eq:Q0}
\end{equation}

\noindent
where $\hat{U}_0$ is a unitary operator that applies only in the chirality space 
and takes the role of the 
``coin'' in the classical random walk. 
Here we focus the case of the Hadamard walk, in which 
$\hat{U}_0 = \hat{I} \otimes \hat{w}_0$, 
where 

\begin{equation}
\hat{w}_0 =   
\frac{1}{\sqrt{2}} \left( \begin{array}{cc} 1 & 1 \\ 1 & -1 \end{array}\right)
\label{eq:U0}
\end{equation}

\noindent
is the Hadamard operator. 
The states of the chirality qubit are expressed 
in the standard basis 

\begin{equation}
|R\rangle =\left( \begin{array}{c} 1 \\ 0 \end{array} \right);  
\ \ \ 
|L\rangle =\left( \begin{array}{c} 0 \\ 1 \end{array} \right), 
\label{eq:standardbasis}
\end{equation}

\noindent
and
$\hat{I}$ 
is the spatial identity operator.
The computational basis state vectors are transformed by the Hadamard operator 
according to

\begin{eqnarray}
\hat{U}_0 |n\rangle|R\rangle & = & 
\frac{1}{\sqrt{2}} 
|n\rangle 
(|R\rangle + |L\rangle) \nonumber \\
\hat{U}_0 |n\rangle|L\rangle & = & 
\frac{1}{\sqrt{2}} 
|n\rangle 
(|R\rangle - |L\rangle). 
\label{U0|ns>}
\end{eqnarray}

\noindent
The translation operator 
$\hat{T}$ 
then 
performs the walker's move according to the chirality state 
such that:

\begin{eqnarray}
\hat{T} |n\rangle|R\rangle & = & |n+1\rangle |R\rangle  \nonumber \\
\hat{T} |n\rangle|L\rangle & = & |n-1\rangle |L\rangle. 
\label{eq:U0|ns>}
\end{eqnarray}

\noindent
These operators yield the following recursion equations for the
amplitudes of the quantum state

\begin{eqnarray}
a_{n,R}(t+1) & = & \frac{1}{\sqrt{2}} [a_{n-1,R}(t) + a_{n-1,L}(t)] \nonumber \\
a_{n,L}(t+1) & = & \frac{1}{\sqrt{2}} [a_{n+1,R}(t) - a_{n+1,L}(t)]. 
\label{a_ns(t)}
\end{eqnarray}

\noindent
For a given initial state, 
the recursion equations provide the probability 
of finding the walker at any site $n$ at time $t$. 
Here we focus on the case that at $t=0$
the walker is located at the origin,
namely,
$a_{n,s}(0) = 0$ 
for all 
$n \neq 0$ 
(and $s = L,R$).

\section{The noisy quantum walk}

The quantum walk, like other quantum computing systems,
may be affected by unitary noise.
This noise is due to fluctuations and drifts in the
parameters of the quantum Hamiltonian of the system. 
The perturbed Hamiltonian is still Hermitian and therefore 
generates time evolution 
operators that  are unitary,
but now include a stochastic part.
Formally, 
one can write a noisy unitary operator as

\begin{equation}
\hat{U} = \hat{U}_0  e^{i \hat{A}},
\label{eq:noisyU}
\end{equation}

\noindent
where 
$\hat{A}$ 
is a stochastic Hermitian operator determined by 
the perturbation 
and 
$\hat{U}_0$ 
is the time evolution operator of the original quantum process 
without the perturbation. 

The Hadamard walk model consists of Hadamard gates 
that act in the chirality space 
at each time step of the walker. 
Here we consider the effect of unitary noise 
in the Hadamard operator on the quantum walker.
To this end we describe a move of the noisy walker at 
time $t$ as:

\begin{equation}
| \psi(t+1) \rangle = \hat{Q}(t) | \psi(t) \rangle.
\label{eq:psi_t_noisy_walk}
\end{equation}

\noindent
where

\begin{equation}
\hat{Q}(t) =  \hat{T} ( \hat{I} \otimes \hat{w}_0 e^{i \hat{a}(t)}).
\label{Q}
\end{equation}

\noindent
The operator 
$\hat{a}(t)$ 
is a stochastic and Hermitian operator 
that applies in the single-qubit chirality space  
at time $t$. 
It can be expanded 
in the basis of the Pauli operators such that

\begin{equation}
\hat{a}(t) = \alpha_1(t) \hat{\sigma}_1 +\alpha_2(t) 
\hat{\sigma}_2+ \alpha_3(t) \hat{\sigma}_3,
\label{eq:a(t)}
\end{equation}

\noindent
where
$\alpha_k(t)$, $k = 1,2,3$ are real stochastic 
variables, 
and the $\sigma$'s are the Pauli 
operators, 
represented in the spin basis as:

\begin{equation}
{\hat{\sigma}}_{1}  =  \left( \begin{array}{cc} 0 & 1 \\ 1 & 0 \end{array} \right); 
\hspace{.5in}
{\hat{\sigma}}_{2}  =  \left( \begin{array}{cc} 0 & -i \\ i & 0 \end{array} \right);
\hspace{.5in}
{\hat{\sigma}}_{3}  =  \left( \begin{array}{cc} 1 & 0 \\ 0 & -1 \end{array} \right). 
\label{eq:hatsigma}
\end{equation}   

\noindent
Although, in principle, the 
$2 \times 2$ identity operator should 
also be included in the expansion 
(\ref{eq:a(t)}), 
it is omitted because it can only change 
the overall phase in the quantum 
walk process, and does not have any 
effect on the walker's measurement probabilities. 

We will now analyze the properties of the noisy 
Hadamard walk according to 
the characteristics of the real stochastic variables $\alpha_k(t)$,
$k=1,2$ and $3$. 
In the analysis we assume that 
there is no correlation between 
different noise components $k$ and 
$k^{\prime} \ne k$
as well as between
different walker steps
at times $t$ and
$t^{\prime} \ne t$, 
namely 

\begin{equation}
\langle \alpha_{k} (t)  \alpha_{k'} (t') \rangle = 
{\delta}_{k,k^{\prime}} {\delta}_{t,t^{\prime}} {\alpha}^2,  
\label{eq:time_uncor}
\end{equation}

\noindent
where 
$\delta_{k,k'}$ 
is the Kronecker delta function.
Furthermore, 
we focus on the isotropic case in which 
$\alpha_1$,
$\alpha_2$
and 
$\alpha_3$
are taken from the same distribution
$p(\alpha)$,
which is unbiased,
namely

\begin{equation}
\langle \alpha_{k} (t) \rangle = 0, \ \ k=1,2,3, 
\label{<alpha>}
\end{equation}

\noindent
with
standard deviation $\alpha$.

To analyze 
the time evolution of the Hadamard walk in the presence 
of unbiased and isotropic unitary noise, 
we have performed computer simulations of the model.  
In these simulations the Hadamard gate is replaced 
by a noisy one, so the walker step at time 
$t$ is given by 
Eq.~(\ref{Q}).  
The Pauli coefficients 
$\alpha_1 (t)$, 
$\alpha_2 (t)$ 
and 
$\alpha_3 (t)$ 
are taken from a Gaussian 
distribution with zero average and a standard deviation $\alpha$. 
The magnitude of the isotropic noise is 
characterized by the values of 
$\alpha$. 
To obtain proper statistics the results were
averaged over a sufficient number of runs. 
For each value of 
$\alpha$, 
we have applied $10,000$ steps of the noisy Hadamard walk for at least 
$200$ runs in the case of 
weak noise 
($\alpha < 0.07$) 
and more than $4,000$ runs in case of strong noise 
($\alpha \geq 0.07$). 
The distribution 
$P_t(n)$
of the walker's position was then examined 
and its moments are calculated. 

The probability to measure the noisy Hadamard walker 
in site $n$ at time $t$ is given by: 

\begin{equation}
\langle P_t(n) \rangle_{\alpha} = \langle {|{a_{n,R}}(t)|}^2 
\rangle_{\alpha} + 
\langle {|{a_{n,L}}(t)|}^2 \rangle_{\alpha}, 
\label{P_n_alpha}
\end{equation}

\noindent
where 
${a_{n,R}}(t)$ 
and  
${a_{n,L}}(t)$ 
are  
amplitudes of 
the walker's state 
$| \psi(t) \rangle$ 
[see Eq.~(\ref{eq:psi_t})].  
Here
$\langle \rangle_{\alpha}$ 
denotes the averaging over the noise,
taken from a Gaussian distribution
with a zero average and 
standard deviation
$\alpha$.

Here we consider only initial conditions for which at $t=0$,
the walker is located at the origin, 
namely
$a_{n,s}(0) = 0$ 
for all 
$n \neq 0$. 
For these initial states it is guaranteed that  
$P_t(n) = 0$ 
for all $|n|>t$. 
Therefore the first and the second moments of the spatial distribution 
at time $t$ 
(which are both averaged over the noise) 
are given by:

\begin{equation}
\langle \overline{{n} (t)} \rangle_{\alpha} = 
\sum_{n = -t}^{t} n \langle P_t(n) \rangle_{\alpha} 
\label{<n_alpha>}
\end{equation}

and 

\begin{equation}
\langle \overline{{n^2} (t)} \rangle_{\alpha} = 
\sum_{n = -t}^{t} n^2 \langle P_t(n) \rangle_{\alpha}. 
\label{<n^2_alpha>}
\end{equation}

\noindent
The standard deviation 
$\sigma_\alpha (t)$ 
of the spatial distribution 
of the noisy Hadamard walk at time $t$ is given by

\begin{equation}
{\sigma}_\alpha^2 (t) 
\equiv 
\langle \overline{{n^2} (t)} \rangle_{\alpha} - 
{\langle \overline{{n} (t)} \rangle_{\alpha}}^2 .
\label{sigma_alpha_t}
\end{equation}

\section{Simulations and results}

To analyze the effect of noise on the Hadamard walk we
have performed direct computer simulations of the system,
using Eq. (\ref{Q}).
The initial condition used in the simulations is

\begin{equation}
|\psi(0) \rangle = \frac{1}{\sqrt{2}}(|0\rangle|R\rangle + i|0\rangle|L\rangle).
\label{eq:psi0_symetric}
\end{equation}

\noindent
For this state the probability
distribution 
$P_t(n)$ 
of the Hadamard
walk 
without noise
turns out to be symmetric around the origin
\cite{Travaglione2002},
namely
$P_t(-n) = P_t(n)$.
In Fig. 
\ref{fig1}
we present the probability distribution 
$P_t(n)$, $n=-t,-t+2,\dots,t$,
for $t=250$ (a), $1000$ (b), and $10,000$ (c) steps.
The distribution exhibits two main peaks, on opposite sides
of the origin, that move away from each other as time evolves.
In the central part it forms a plateau, in which
$P_t(n) \sim 1/t$.
It also exhibits wild oscillations under the envelope, that
decays towards the origin and rises towards the two peaks.
This distribution is clearly very different from the Gaussian
distribution obtained for the classical random walk.
The formation of 
$P_t(n)$
can be understood as a result of constructive interference
in the front of the propagating wave function, while the
center is dominated by destructive interference.

When unitary noise is added, it perturbs the structure 
produced by constructive and destructive interference.
The distribution 
$P_t(n)$ for a weak noise level
($\alpha = 0.025$)
is shown in  
Fig.~\ref{fig2}
for $t=250$ (a), $1000$ (b), and $10,000$ (c) steps.
The effects of the competition 
between the behavior of the original Hadamard walk 
and the unitary noise are clearly shown. 
First, alongside the oscillating wings on the edges, an
incoherent component is generated around the center
[Fig.~\ref{fig2}(a)]. 
Gradually, the 
central peak increases absorbing weight from
the two oscillating wings  
[Fig.~\ref{fig2}(b)], 
which later disappear completely, 
[Fig. \ref{fig2}(c)]. 
However, the probability distribution does not converge to a Gaussian 
even at much longer times,
since at any time step the quantum effects continue to be dominant. 
The incoherent component becomes more dominant as time evolves, 
due to the accumulation of noise effects.
However, the coherent features of the quantum walk appear in every
time step and the probability distribution does not seem to 
converge to a Gaussian.                                           

In Fig.~\ref{fig3} 
we present the probability distribution
$P_t(n)$
for the case of strong noise
($\alpha = 0.8$)
at $t=1000$.
At this noise level 
the quantum effects are completely suppressed by the noise.
The oscillations are 
smoothed and the 
probability distribution converges towards 
a Gaussian distribution with a zero mean, that 
expands with time. 
It thus approaches the behavior of the classical random
walk.

In 
Fig.~\ref{fig4} 
we present
the time dependence of the standard 
deviation 
$\sigma_\alpha (t)$ 
of
$P_t(n)$
for the noisy Hadamard walk 
with symmetric, unbiased noise.
The standard deviation
$\sigma_\alpha (t)$,
given by 
Eq.~(\ref{sigma_alpha_t}),
is shown for 
different noise levels
$\alpha$.
The top curve is for the noiseless quantum walk,
the next four curves, from top to bottom, are for
$\alpha = 0.025$, $0.05$, $0.1$ and $0.2$, and the
last curve at the bottom is for the classical random walk.
The initial state used in these simulations is

\begin{equation}
|\psi(0) \rangle = |0\rangle |R\rangle.
\label{psi0_0+}
\end{equation}

\noindent
We observe that as the noise level $\alpha$
increases, the standard deviation curves
for the noisy quantum walk 
move down from the top curve of the noiseless quantum walk
and approach the bottom curve describing the classical 
random walk.

The standard deviation
$\sigma_0(t)$ 
for the noiseless quantum walk 
as a function of time $t$ 
takes the form 
\cite{Ambainis01}

\begin{equation}
\sigma_0 (t) = q t. 
\label{sigma0}
\end{equation}

\noindent
From our numerical simulations we obtain that
the coefficient 
$q = 0.4505 \pm 0.0005$, 
which is in perfect agreement with the analytical
results of Ref.
\cite{Ambainis01}.
The standard deviation for the classical random walk is
given by

\begin{equation}
\sigma_{\mbox{classical}}(t) = \sqrt{t}.
\label{sigma_classical}
\end{equation}

\noindent
As long as the noise level is not
too high, the noisy Hadamard walk shows 
the typical behavior of a quantum walk for
short times. 
However, beyond some crossover time 
the motion of the noisy quantum walk 
acquires diffusive features in the sense
that $\sigma_{\alpha}(t)$
behaves like the classical random walk.
However, the shape of $P_t(n)$
does not approach a Gaussian and maintains
a far reaching tail.

In order to evaluate the crossover time,
and examine its dependence on $\alpha$,
we will use the short-time and long-time
limits of
$\sigma_\alpha (t)$.
For short times
$\sigma_\alpha (t)$
increases
linearly with $t$, 
according to Eq.~(\ref{sigma0}). 
Then, gradually the accumulation of the noise 
modifies the curve until 
in the long time limit
it forms a square-root shape.  
The long-time tail of 
$\sigma_\alpha (t)$
is well fitted by
$\sigma_\alpha (t) = K(\alpha) \sqrt{t} + C(\alpha)$, 
and the functions
$K(\alpha)$ 
and 
$C(\alpha)$ 
can be obtained.
It turns out that for long enough times
$C(\alpha)$ 
can be neglected,
giving rise to

\begin{equation}
\sigma_\alpha (t) = K(\alpha) \sqrt{t}. 
\label{eq:sigma_alpha}
\end{equation}

\noindent
The function 
$K(\alpha)$ 
is then obtained
by linear regression of the points 
$[\sqrt{t},\sigma_\alpha(t)]$,
for each value of $\alpha$. 
The function
$K(\alpha)$ 
for a relevant range of
$\alpha$ values is shown in
Fig.~\ref{fig5}.
It turns out to be 
a monotonically decreasing function,
that in
the limit of small $\alpha$ 
takes the form
$K(\alpha) \sim 1/\alpha$.
In the limit of very large
$\alpha$,
$K(\alpha) \rightarrow 1$,
and the noisy quantum walk 
coincides with the classical random walk,
for which $K=1$ (dashed line).
The monotonically decreasing behavior of
$K(\alpha)$
means that as the noise level increases, the
broadening of the distribution
$P_t(n)$ in the long time limit becomes slower.
This is due to an earlier suppression of the quantum 
coherence, that broadens the distribution much faster
than the noise.

We will now consider the crossover time $T_2(\alpha)$, 
which is based on the behavior of
$\sigma_{\alpha}(t)$.
It is defined as the time at which
the long time asymptote of the 
noisy Hadamard walk 
[given by Eq.~(\ref{eq:sigma_alpha})]
intersects with the short time asymptote,
namely the ideal noiseless linear curve 
[Eq. (\ref{sigma0})]. 
This crossover time is given by

\begin{equation}
K(\alpha) \sqrt{T_2(\alpha)} = q T_2(\alpha),
\end{equation}

\noindent
and therefore

\begin{equation}
T_2(\alpha) = {\left[ \frac{K(\alpha)}{q} \right]}^2.
\label{crossover_time2}
\end{equation}

\noindent
The crossover time 
$T_2(\alpha)$, obtained from
Eq. 
(\ref{crossover_time2}),
is shown
in 
Fig.~\ref{fig6} 
as a function of 
$\alpha$,
on a log-log scale. 
It is found that

\begin{equation}
T_2 = c_2 \  {\alpha}^{-\eta},
\label{T2_eta}
\end{equation}  

\noindent
where 
$c_2 = 0.62 \pm 0.01$
and
$\eta = 2.05 \pm 0.08$. 
The crossover time can be interpreted as the
time required for the noise to scramble the
phases and thus eliminate the structure of
the constructive and destructive interference
that characterize the Hadamard walk.
This time can be estimated by

\begin{equation}
\alpha \sqrt{T_2} \sim 1
\label{eq:argt2}
\end{equation}

\noindent
namely,

\begin{equation}
T_2 \sim \frac{1}{{\alpha}^2},
\label{T2_vs_alpha}
\end{equation}

\noindent
in agreement with the simulation results.
Unlike the classical random walk that produces a symmetric
probability distribution for a particle that starts from
the origin, the Hadamard walk exhibits an inherent asymmetry. 
For example, the initial state
of Eq.
(\ref{psi0_0+})
produces a distribution
$P_t(n)$
with much more weight on the right vs.
the left hand side.
Moreover, the average position of the walker moves
to the right at a constant speed,
namely

\begin{equation}
\overline{n(t)} = v t.
\label{eq:nbarvt}
\end{equation}

\noindent
where $v = 0.293 \pm 0.003$, 
in perfect agreement with the analytical results of
Ref.
\cite{Ambainis01}.
In Fig.~\ref{fig7} 
we show the evolution in time of 
the average position
$\langle \overline{n(t)} \rangle_{\alpha}$
[given by  Eq.~(\ref{<n_alpha>})] 
for the noisy Hadamard walk 
with different values of the noise 
$\alpha$.
The dashed linear line at the top is the result for
the noiseless Hadamard walk,
given by
Eq.~(\ref{eq:nbarvt}).
Below it, from top to bottom are the simulation results for
$\alpha = 0.025, 0.03, 0.04, 0.07$ and $0.1$.
For short times, the average
$\langle \overline{n(t)} \rangle_{\alpha}$
of the noisy Hadamard walk tends to 
follow the straight line of 
Eq.~(\ref{eq:nbarvt}),
up to some crossover time,
$T_1(\alpha)$, 
at which it saturates and approaches a constant value:

\begin{equation}
\langle \overline{n(t)} \rangle_{\alpha} \rightarrow n_\alpha.
\label{n_alpha}
\end{equation}

\noindent
The asymptotic value, $n_\alpha$,
decreases as the noise level
$\alpha$
is increased.
This is due to the fact that at higher noise levels, the buildup of the asymmetric
pattern, which is a quantum effect, is suppressed more quickly.
We define the crossover time
$T_1(\alpha)$ 
as the time $t$ at which the line of 
Eq.~(\ref{eq:nbarvt}),
describing the noiseless Hadamard walk,
intersects the
asymptotic horizontal line
$\langle \overline{n(t)} \rangle_{\alpha} = n_\alpha$,
namely
$v T_1 = n_\alpha$.
Therefore

\begin{equation}
T_1 = \frac{n_\alpha}{v}.
\label{crossover_time1}
\end{equation}

\noindent
In 
Fig.~\ref{fig8} 
we present the crossover time
$T_1(\alpha)$
vs. $\alpha$
on a log-log scale.
It is well fitted by a power-law function of the form

\begin{equation}
T_1 = c_1 \  {\alpha}^{-\rho},
\label{T1_rho}
\end{equation} 

\noindent
where 
$c_1 = 0.20 \pm 0.01$
and
$\rho = 2.06 \pm 0.08$,
which is consistent with the expected asymptotic result
of $\rho = 2$, based on an argument
similar to
Eq.~(\ref{eq:argt2}).  
We thus find that both 
$T_1(\alpha)$
and 
$T_2(\alpha)$
exhibit similar dependence on the
noise level $\alpha$, up to a constant factor.

We will now discuss the connection between our results for the
effect of unitary noise on the quantum walk and earlier results
on decoherence effects
\cite{Brun2002,Kendon2002,Brun2003,Kendon2003,Kendon2003b}. 
Decoherence can be described by an additional
quantum operation that applies on the chirality qubit at each step
of the quantum walk. 
In a particular choice of the noise model,
this operation is given by 
the completely positive map that consists of the operators
\cite{Brun2003}

\begin{eqnarray}
\hat A_0 = \sqrt{p} | R \rangle \langle R | \nonumber \\
\hat A_1 = \sqrt{p} | L \rangle \langle L | \nonumber \\
\hat A_2 = \sqrt{1-p} \hat I, 
\end{eqnarray}

\noindent
where $\hat I$ is the identity operator.
This set of operators satisfies

\begin{equation}
\sum_{n=0}^2 \hat A_n^{\dagger} \hat A_n = \hat I.
\end{equation}

\noindent
It can be interpreted as a measurement of the chirality qubit
that is performed with probability $p$.
Analytical calculations show that in the long time limit, the
first moment of the decohered quantum walk, starting in the state
$| 0 \rangle | R \rangle$,
saturates and approaches the value:

\begin{equation}
\langle \overline{n(t)} \rangle_{p} \rightarrow n_p = \frac{(1-p)^2}{p(2-p)}.
\label{n_p}
\end{equation}

\noindent
For weak decoherence, or $p \ll 1$, 
$n_p \sim 1/p$,
which resembles our result
$n_{\alpha} \sim 1/\alpha^2$,
if $p$ is replaced by $\alpha^2$.
This is a sensible connection since $\alpha$
represents some matrix elements that multiply
the amplitudes while $p$ is a probability.

The results of Ref.
\cite{Brun2003}
for the second moment
of the spatial distribution of
the decohered quantum walk 
can be expressed by

\begin{equation}
\sigma_p (t) = K(p) \sqrt{t}, 
\label{eq:sigma_p}
\end{equation}

\noindent
where

\begin{equation}
K(p) = \sqrt{ 1 + \frac{2(1-p)^2}{p(2-p)} }.
\end{equation}

\noindent
Like $K(\alpha)$, this is a monotonically decreasing function,
that in the limit of small $p$ takes the form $K(p) \sim 1/\sqrt{p}$,
while 
$K(p) \rightarrow 1$
for 
$p \rightarrow 1$. 
Therefore, associating $p$ with $\alpha^2$, one obtains the same
scaling behavior of the noise effects in the two noise
models for both the first and second moments of the distribution. 

\section{Summary}

We have studied the effect of unitary noise on the discrete 
one dimensional quantum walk. 
We have shown that when the quantum walk is exposed to unitary noise 
it exhibits a crossover from the quantum behavior for short times 
to classical-like behavior for long times. 
For times shorter than the crossover time,
the standard deviation 
$\sigma_{\alpha}(t)$ 
of the spatial distribution of the noisy Hadamard walk, 
scales linearly with $t$, 
like the noiseless Hadamard walk. 
Beyond the crossover time, 
$\sigma_{\alpha}(t)$ 
scales like 
$\sqrt{t}$, 
namely it acquires diffusive behavior, like
the classical random walk. 
The crossover time was also characterized using the average position
of the random walker,
namely the first moment, 
$\langle \overline{n(t)} \rangle_{\alpha}$, 
of the spatial distribution,
which scales like $t$ for short times and saturates to a constant
value for long times. 
In both cases the crossover time was found to scale as
$\alpha^{-2}$
where $\alpha$
is the standard deviation of the random noise.

\acknowledgments
OB thanks the Center for Quantum Computer Technology and the
Department of Physics at the University of Queensland for hospitality
during a sabbatical leave when this work was initiated.  
The work at the Hebrew University was supported
by the EU Grant No.~IST-1999-11234.


\newpage
\clearpage

\begin{figure}
\caption{
The probability distribution $P_t(n)$
of the noiseless Hadamard walk 
on the one dimensional lattice
at time $t$, vs. the spatial coordinate $n$.
This probability distribution is given for 
different times: $t = 250$(a), 
$t=1000$ (b) and $t=10000$ (c).
The initial state of the quantum walk is 
given by 
Eq.~(\ref{eq:psi0_symetric}).
}
\label{fig1}
\end{figure}

\begin{figure}
\caption{
The probability distribution 
$\langle P_t(n) \rangle_{\alpha}$ 
of the noisy Hadamard walk 
at low noise level $\alpha = 0.025$, 
as a function of $n$ for 
$t = 250$ (a), $t = 1000$ (b) and $t = 10000$ (c).
The initial state of the walk is given by 
Eq.~(\ref{eq:psi0_symetric}).
}
\label{fig2}
\end{figure}

\begin{figure}
\caption{
The probability distribution 
$\langle P_t(n) \rangle_{\alpha}$ 
of the noisy Hadamard walk
for high noise level 
$\alpha = 0.8$, as a function of $n$, 
for $t = 1000$.
The initial state of the walk is given by 
Eq.~(\ref{eq:psi0_symetric}).
}
\label{fig3}
\end{figure}

\begin{figure}
\caption{
The standard deviation 
$\sigma_\alpha(t)$ 
for the noisy Hadamard walk 
as a function of the time $t$,
for several noise levels $\alpha$.
The top curve (dashed line) is for the 
noiseless Hadamard walk. 
Below it (from top to bottom) are the results for the noisy Hadamard 
walk with $\alpha = 0.025, 0.05, 0.1$ 
and  $0.2$, respectively (solid lines). 
The bottom curve (dashed line) is for
the classical random 
walk.  
Each data point was averaged over 
$200$ runs for weak noise 
($\alpha < 0.07$) 
and over 
$4000$ runs in case of strong noise ($\alpha \geq 0.07$).
}
\label{fig4}
\end{figure}

\begin{figure}
\caption{
The coefficient $K(\alpha)$ 
as a function of 
the noise level $\alpha$.
In the limit of small $\alpha$ it 
takes the form
$K(\alpha) \sim 1/\alpha$.
In the limit of very large
$\alpha$,
$K(\alpha) \rightarrow 1$,
namely, it coincides with 
the result of the classical random walk.
}
\label{fig5}
\end{figure}

\begin{figure}
\caption{
The crossover time $T_2$ 
as a function of 
the noise level $\alpha$, on a log-log scale. 
It follows a power-law behavior
$T_2 \sim {\alpha}^{-\eta}$
where $\eta = 2.05 \pm 0.08$.
}
\label{fig6}
\end{figure}

\begin{figure}
\caption{
The average position
$\langle \overline{n(t)} \rangle_{\alpha}$ 
of noisy Hadamard walk 
as a function of the time $t$.
The top curve is for the 
noiseless Hadamard walk and
below it, from top to bottom are the curves for
$\alpha = 0.025, 0.03, 0.04, 0.07$ and $0.1$. 
}
\label{fig7}
\end{figure}

\begin{figure}
\caption{
The crossover time $T_1$ as a function of 
$\alpha$, on a log-log scale. 
It follows a power-law of the form
$T_1 \sim {\alpha}^{-\rho}$
where
$\rho = 2.06 \pm 0.08$
}
\label{fig8}

\end{figure}

\newpage
\clearpage
\includegraphics[width=17cm]{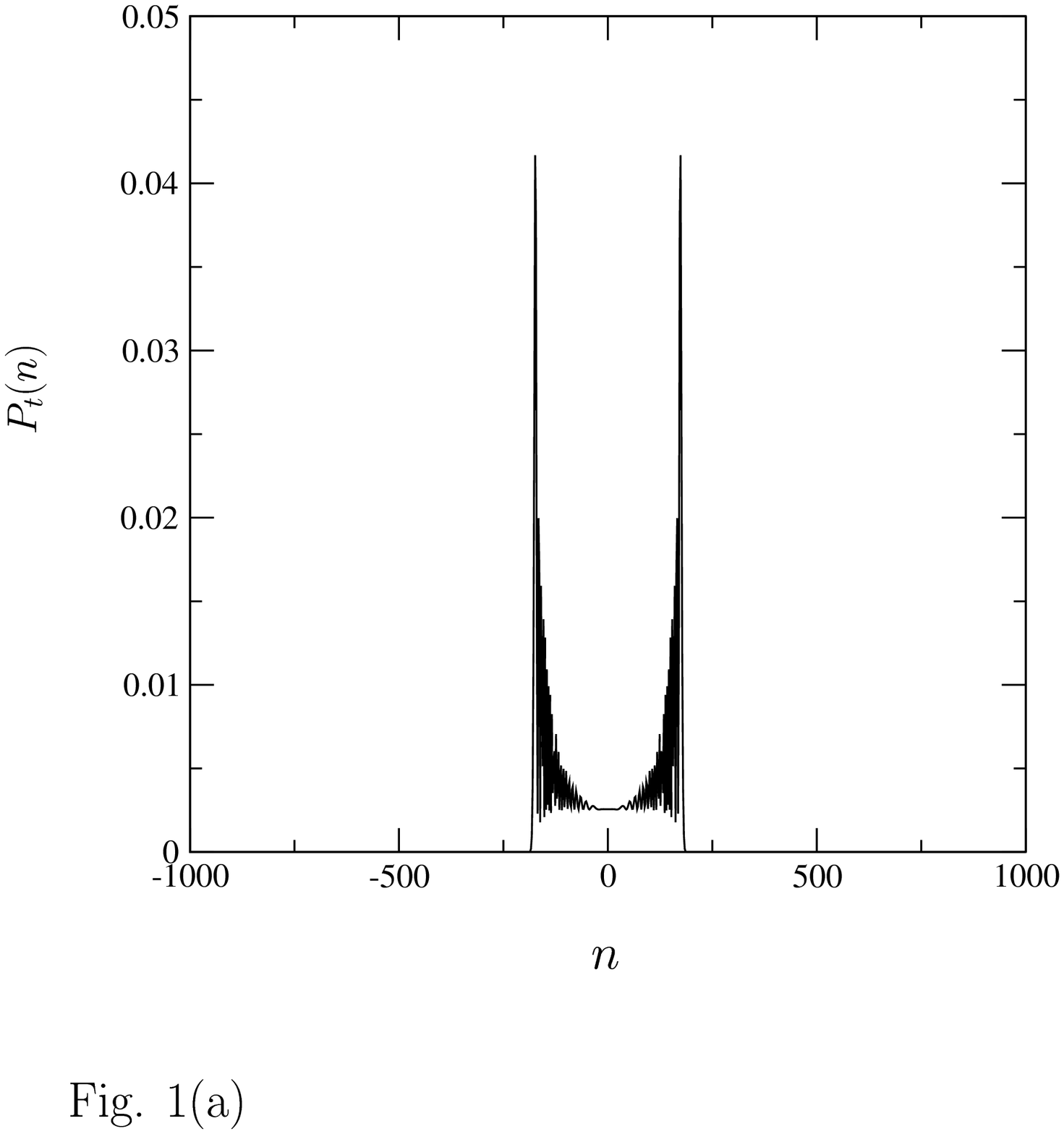}
\newpage
\includegraphics[width=17cm]{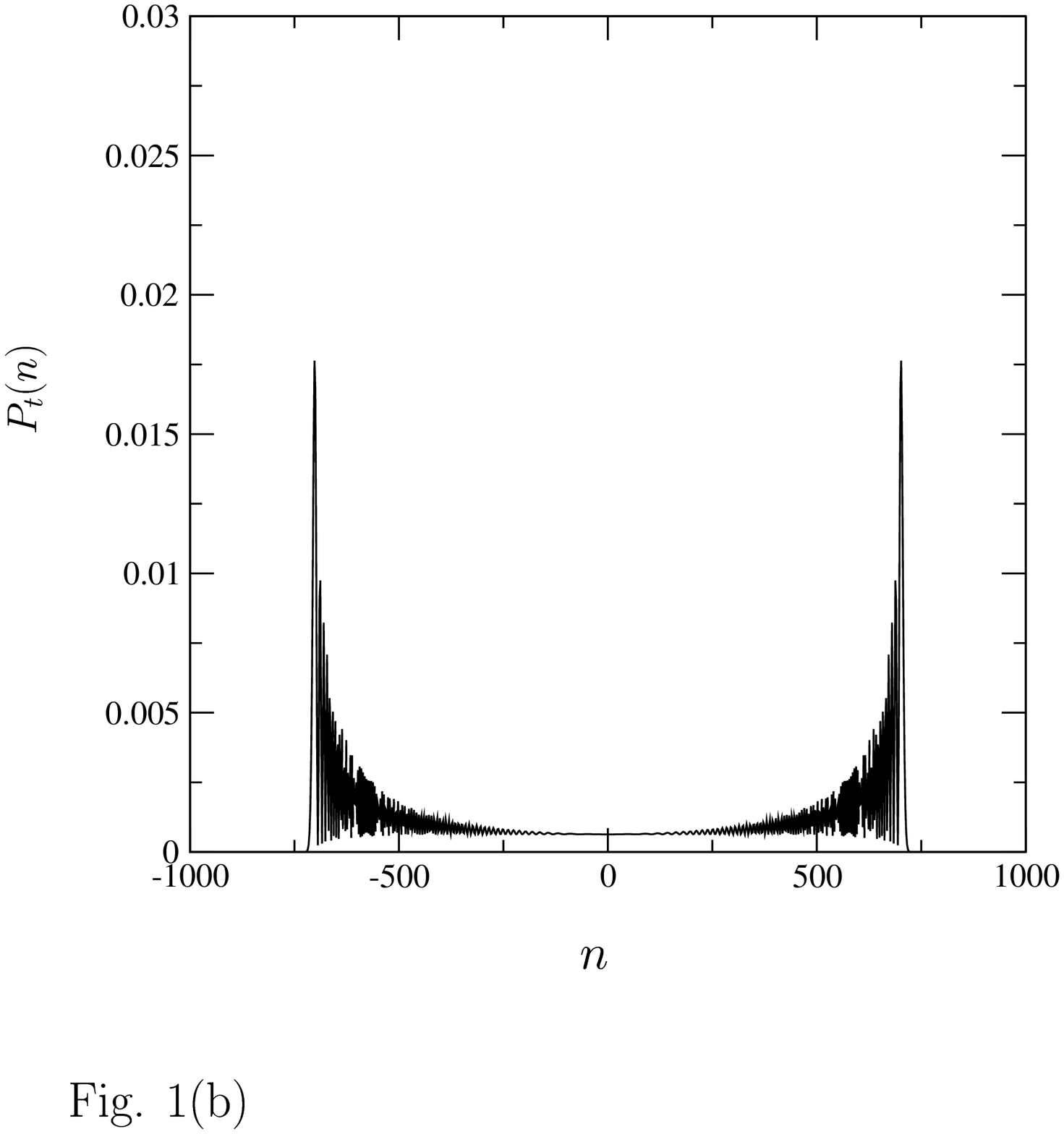}
\newpage
\includegraphics[width=17cm]{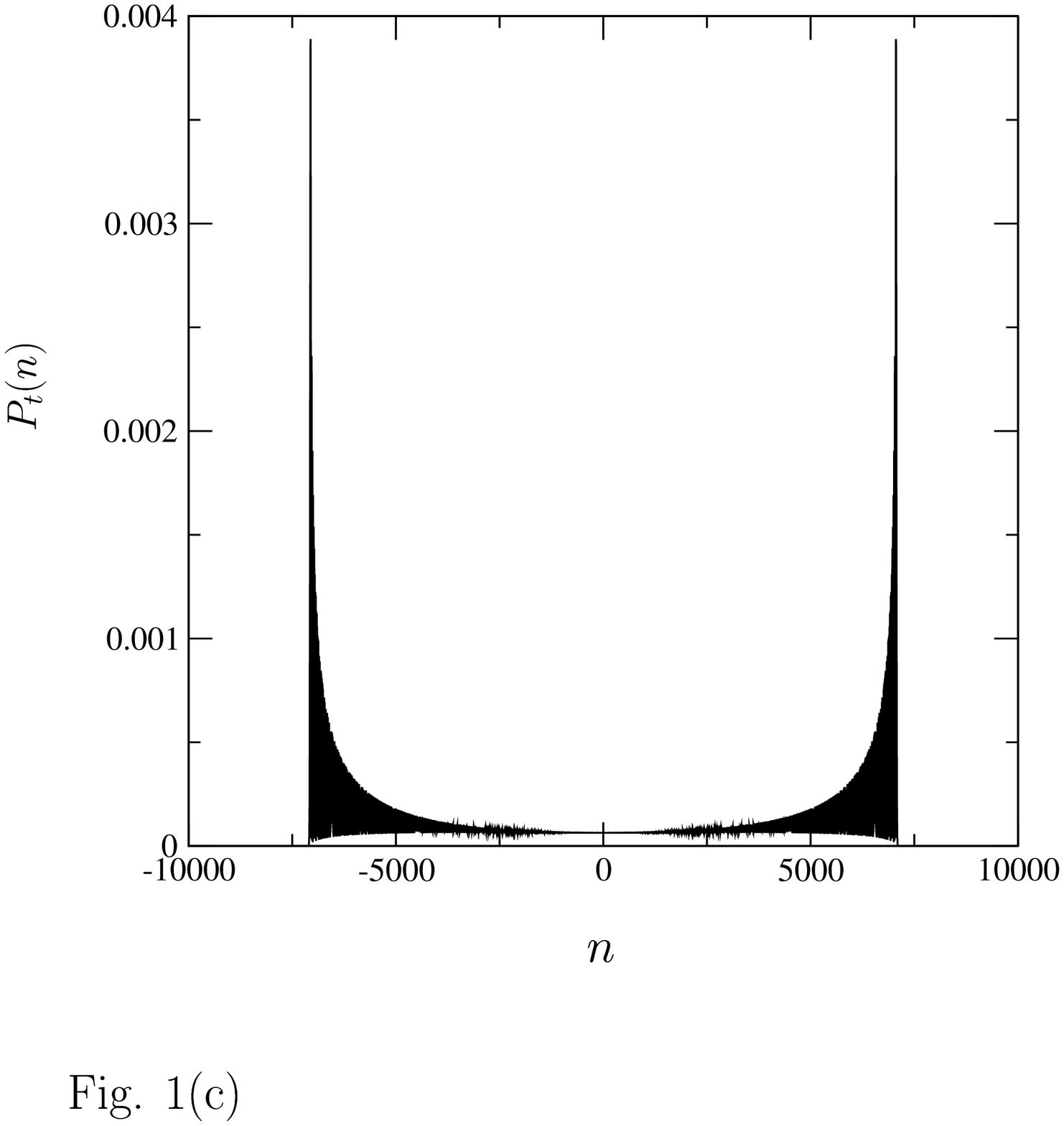}
\newpage
\includegraphics[width=17cm]{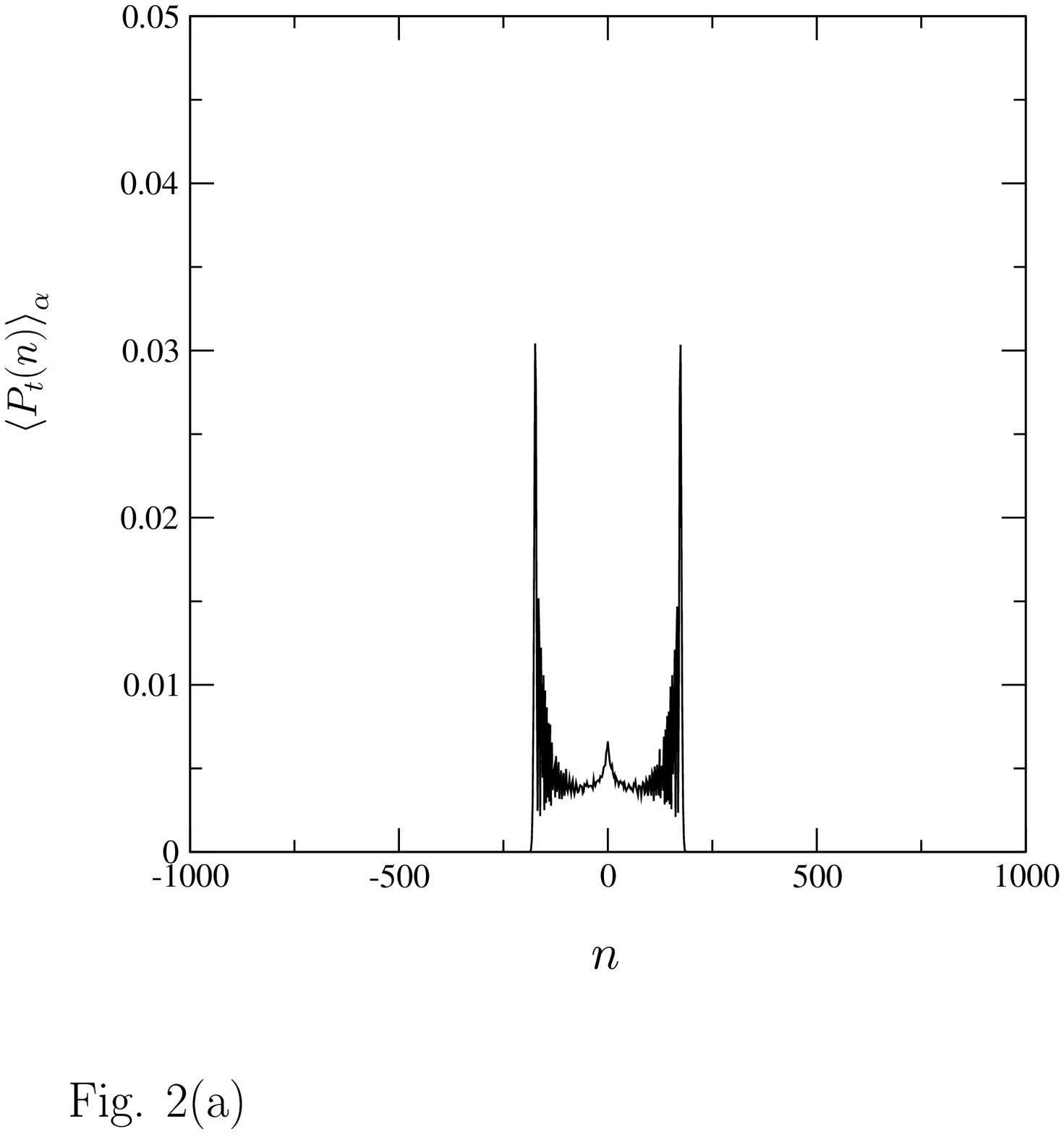}
\newpage
\includegraphics[width=17cm]{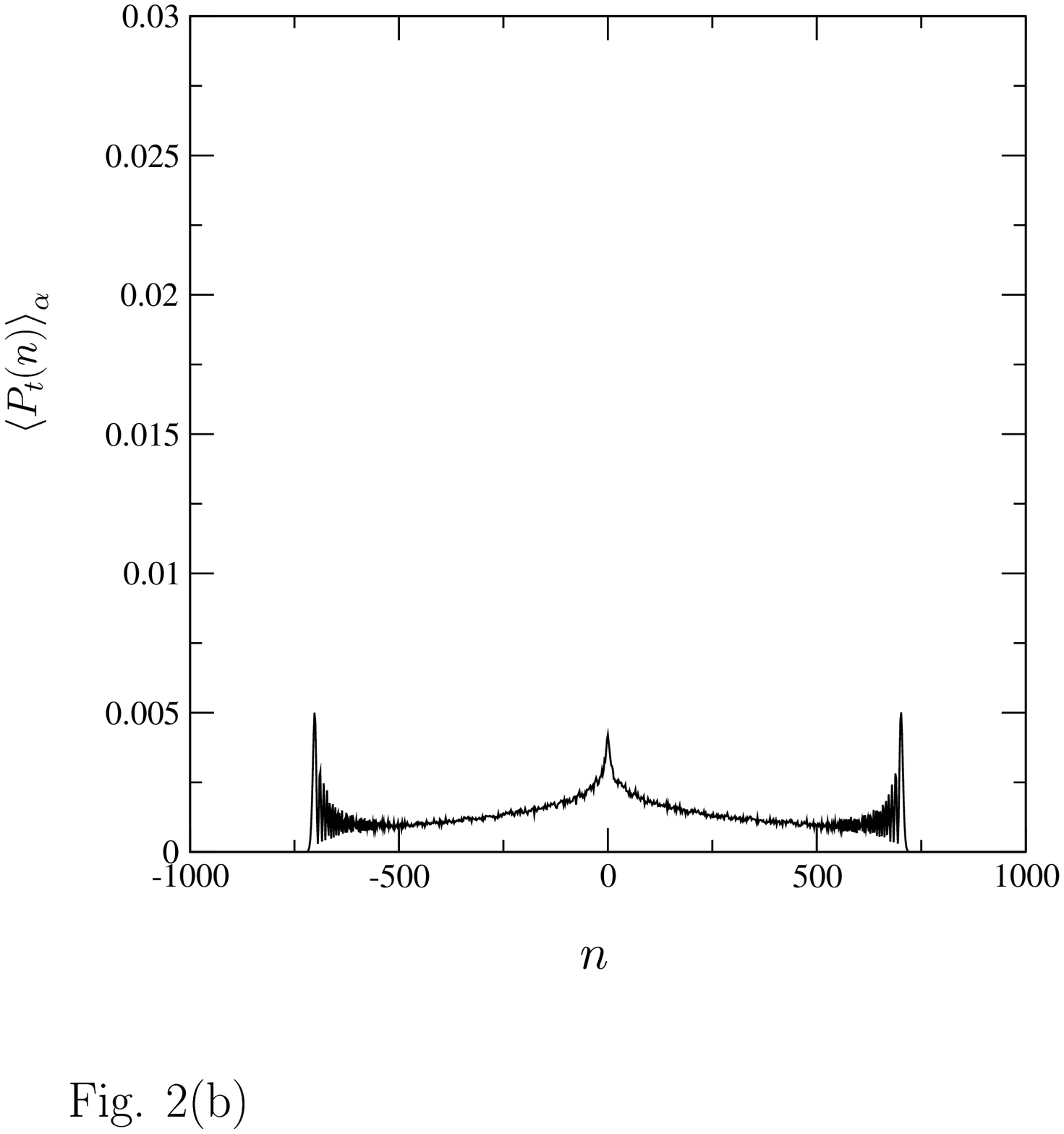}
\newpage
\includegraphics[width=17cm]{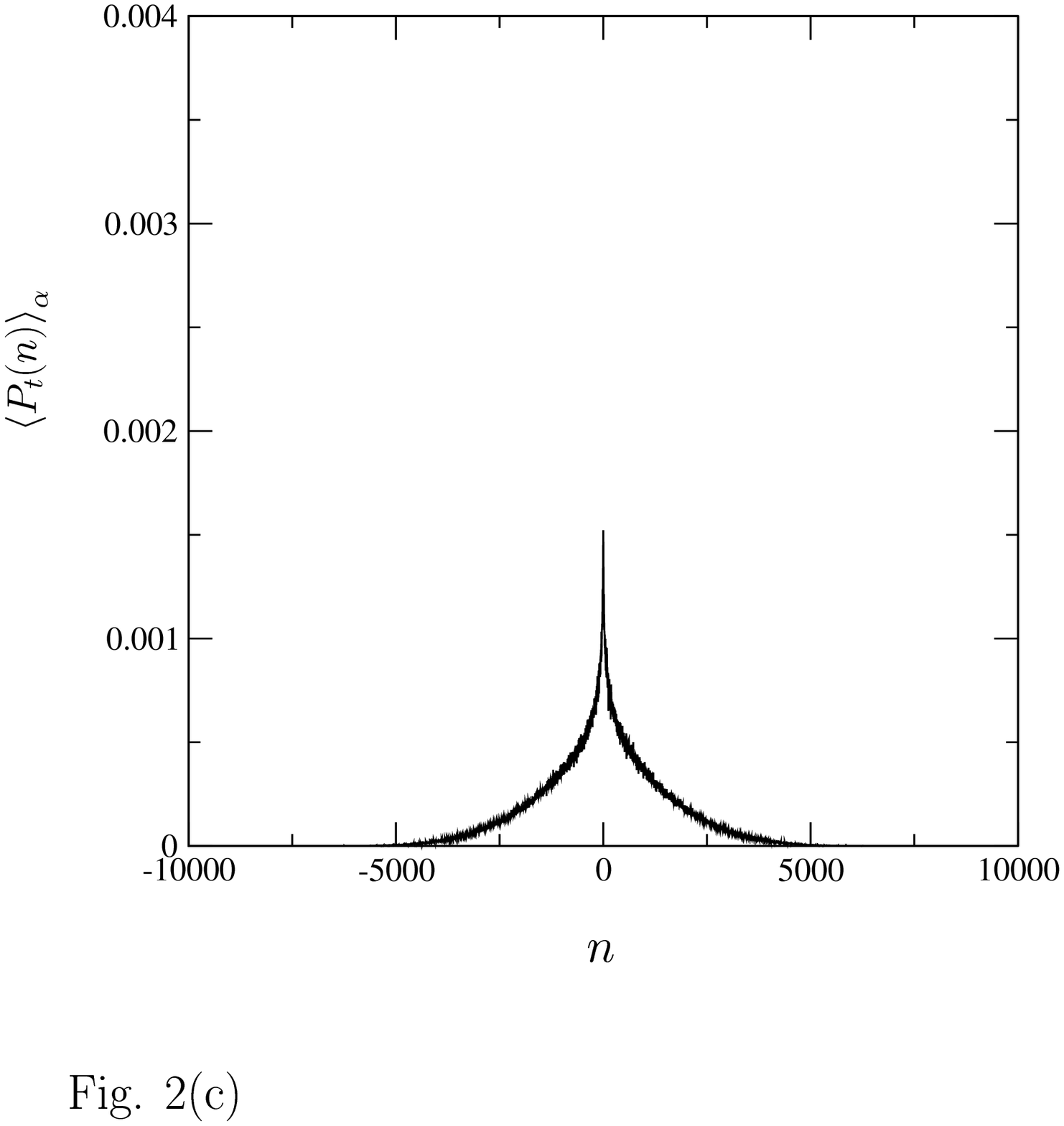}
\newpage
\includegraphics[width=17cm]{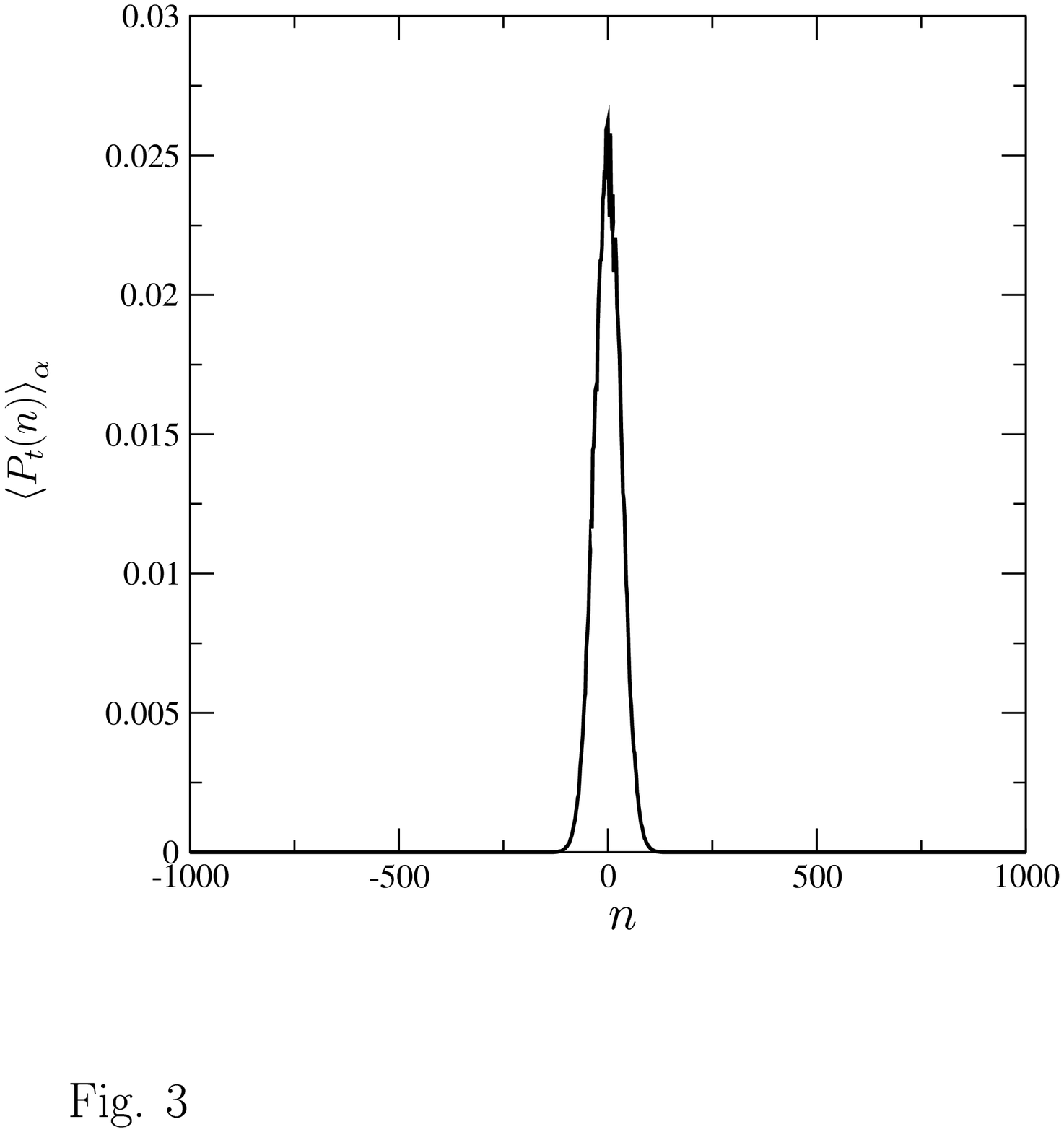}
\newpage
\includegraphics[width=17cm]{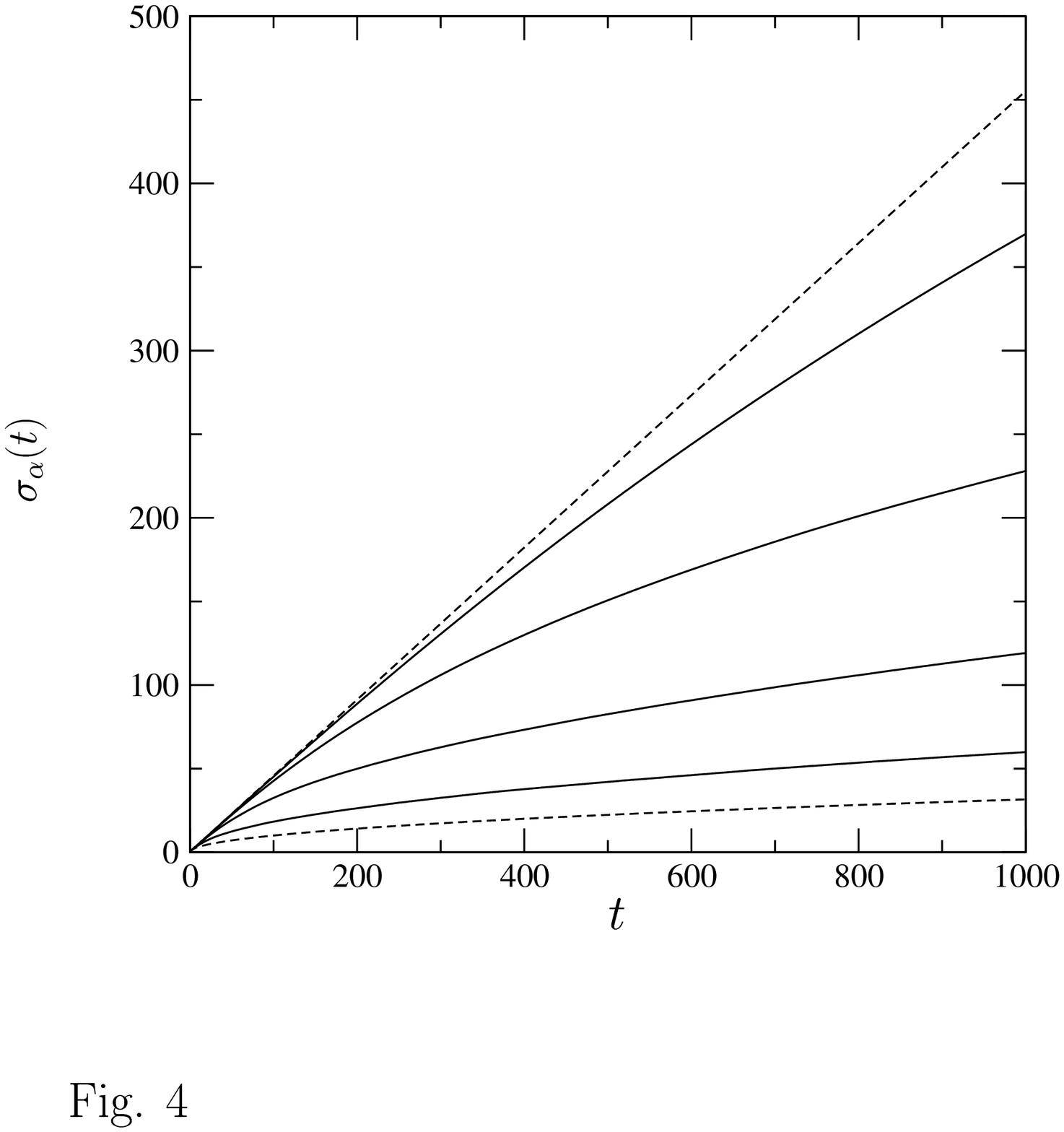}
\newpage
\includegraphics[width=17cm]{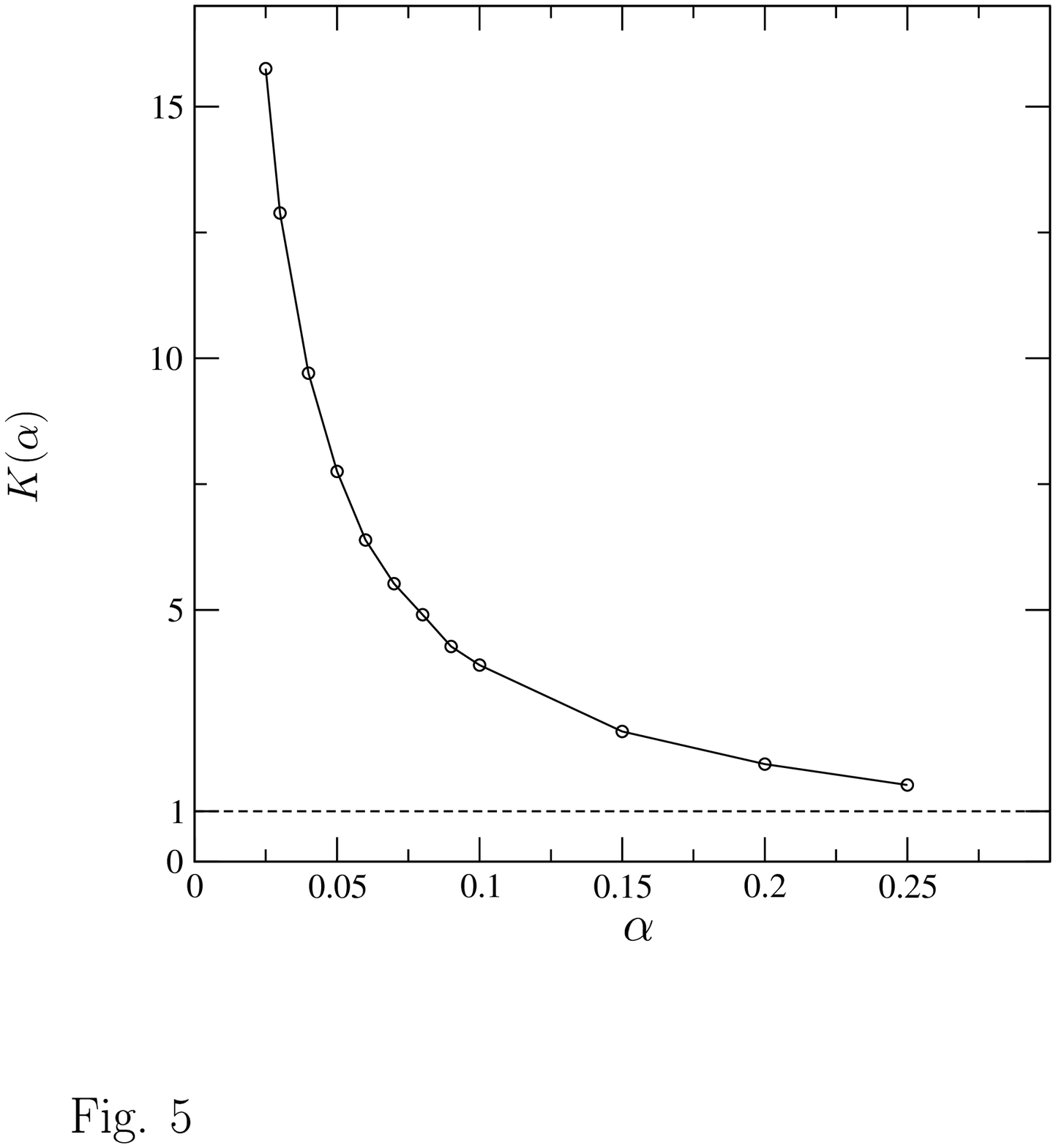}
\newpage
\includegraphics[width=17cm]{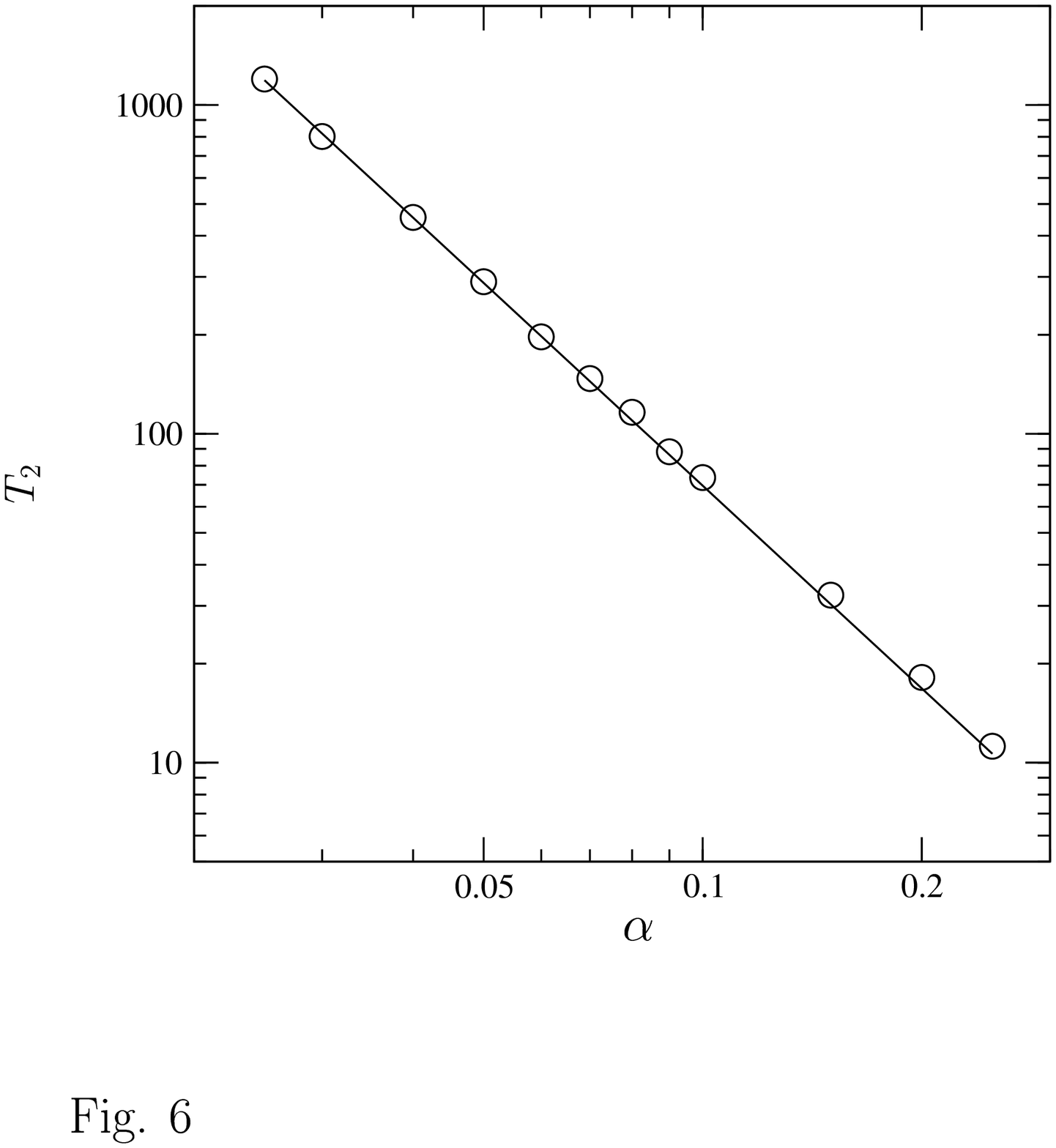}
\newpage
\includegraphics[width=17cm]{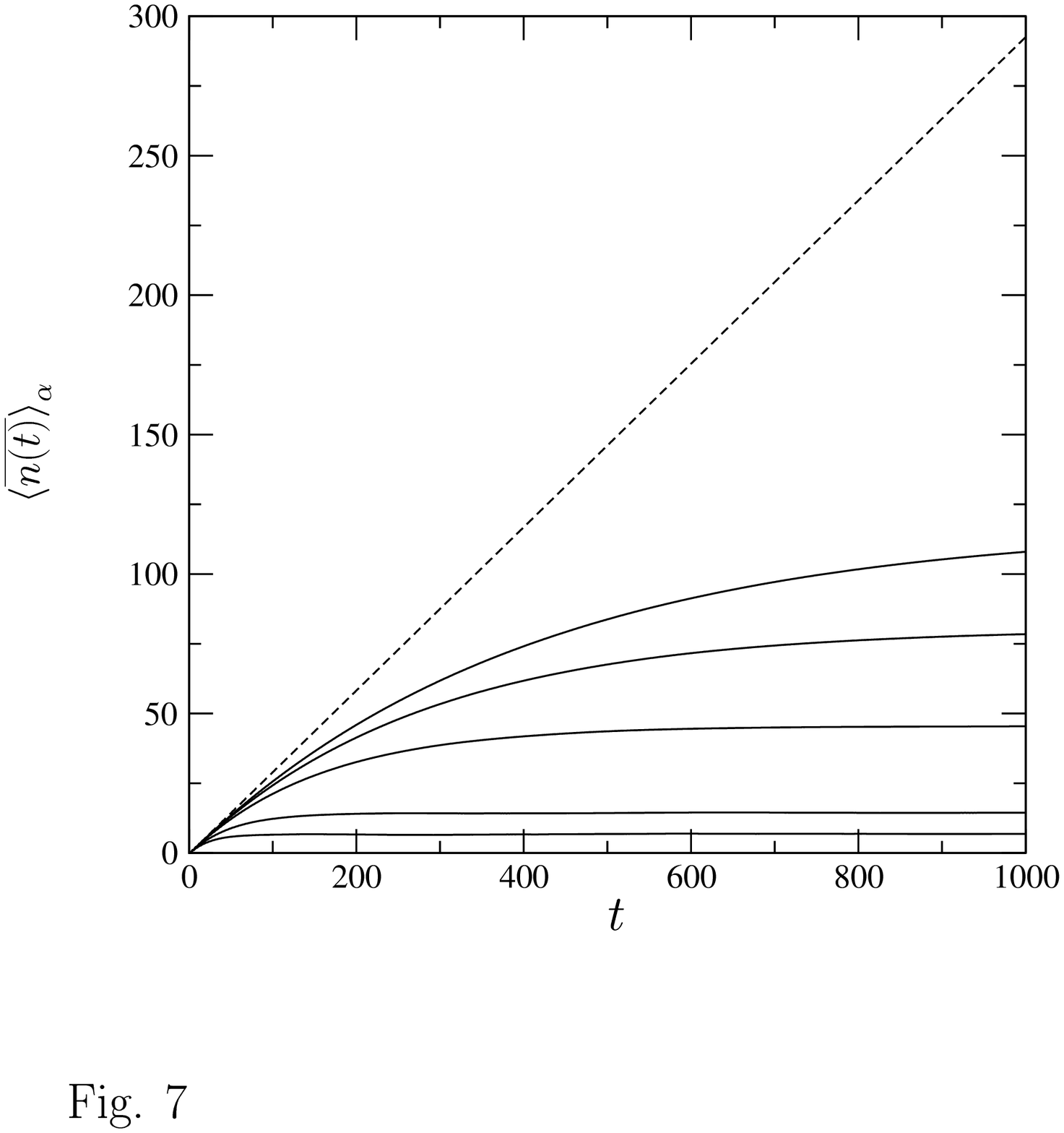}
\newpage
\includegraphics[width=17cm]{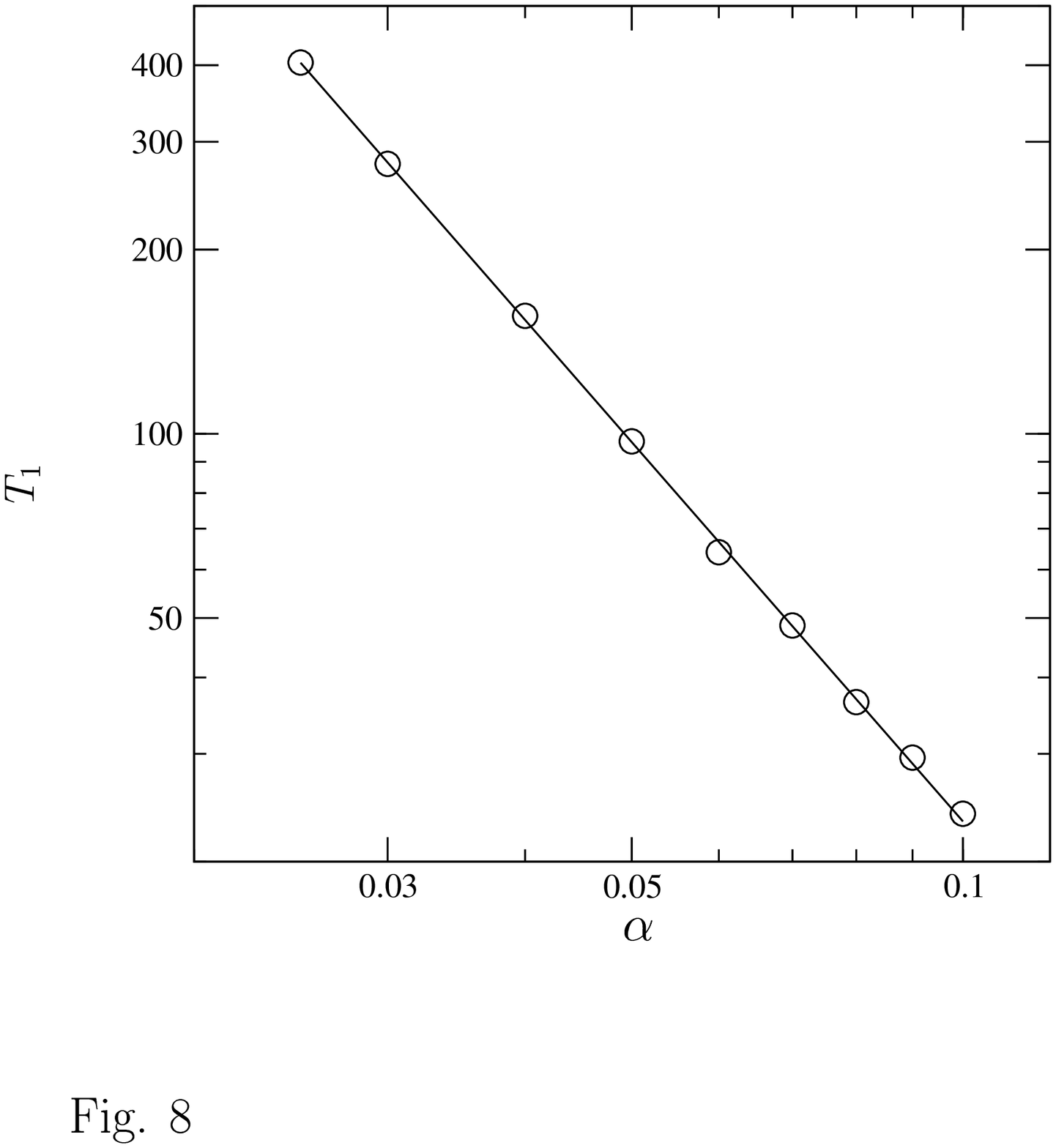}

\end{document}